\documentclass[11pt]{article}    %% 
\usepackage{axodraw,mathrsfs}
\makeatletter

\@addtoreset{equation}{section}
\makeatother

\def\vec#1{\mbox{\boldmath $#1$}}
\def\eqn#1{Eq.\ (\ref{#1})}

\def\gauss{\,{\rm G}}

\def\omit#1{_{\!\rlap{$\scriptscriptstyle \backslash$}
{\scriptscriptstyle #1}}}

\renewcommand{\bar}{\overline}

\title{\centerline{\normalsize SINP/TNP/02-34 \hfill hep-ph/0212118}
\bf Neutrinos and magnetic fields~: a short review} 

\author{
\bf Kaushik Bhattacharya and Palash
B. Pal\thanks{kaushikb@theory.saha.ernet.in, 
pbpal@theory.saha.ernet.in}\\  
\normalsize Saha Institute of Nuclear Physics, 1/AF Bidhan-Nagar, 
Calcutta 700064, India}

\date{November 2002}

\begin{document}

\maketitle 

%%%%%%%%%%%%%%%%%%%%%%%%% 
\begin{abstract}\noindent\small
Neutrinos have no electric charge, but a magnetic field can indirectly
affect neutrino properties and interactions through its effect on
charged particles.  After a brief field-theoretic discussion of
charged particles in magnetic fields, we discuss two broad kinds of
magnetic field effects on neutrinos.  First, effects which come
through virtual charged particles and alter neutrino properties.
Second, effects which alter neutrino interactions through charged
particles in the initial or final state.  We end with some discussion
about possible physical implications of these effects.

\end{abstract} 
%%%%%%%%%%%%%%%%%%%%%%%%% 

%%%%%%%%%%%%%%%%%%%%%%%%%
\section{Motivation}\label{mo} 
%%%%%%%%%%%%%%%%%%%%%%%%%
Neutrinos have no electric charge.  So they do not have any direct
coupling to photons in any renormalizable quantum field theory.  The
standard Dirac contribution to the magnetic moment, which comes from
the vector coupling of a fermion to the photon, is therefore absent
for the neutrino.  In the standard model of electroweak interactions,
the neutrinos cannot have any anomalous magnetic moment either.  The
reason is simple: anomalous magnetic moment comes from
chirality-flipping interactions $\bar\psi\sigma_{\mu\nu}\psi
F^{\mu\nu}$, and neutrinos cannot have such interactions because there
are no right-chiral neutrinos in the standard model.  The bottom line
is: neutrinos do not interact with the magnetic field at all in the
standard model.

Why then should we discuss the relation between neutrinos and magnetic
fields?  There are several reasons, which will be discussed in the
rest of this section.

\begin{figure}[btp]
\begin{center}
%
% photon coupled to electron
%
\begin{picture}(180,120)(-90,-35)
\Text(0,-30)[ct]{\large\bf (a)}
\ArrowLine(80,0)(40,0)
\Text(60,-10)[c]{$\nu$}
\Photon(40,0)(-40,0)37
\Text(0,-10)[c]{$W$}
\ArrowLine(-40,0)(-80,0)
\Text(-60,-10)[c]{$\nu$}
\ArrowArc(0,0)(40,0,90)
\Text(35,40)[l]{$\ell$}
\ArrowArc(0,0)(40,90,180)
\Text(-35,40)[r]{$\ell$}
\Photon(0,40)(0,80){2}{4}
\end{picture}
%
% photon coupled to W
%
\begin{picture}(180,120)(-90,-35)
\Text(0,-30)[ct]{\large\bf (b)}
\ArrowLine(80,0)(40,0)
\Text(60,-10)[c]{$\nu$}
\ArrowLine(40,0)(-40,0)
\Text(0,-10)[c]{$\ell$}
\ArrowLine(-40,0)(-80,0)
\Text(-60,-10)[c]{$\nu$}
\PhotonArc(0,0)(40,0,90){3}{6}
\Text(43,40)[r]{$W$}
\PhotonArc(0,0)(40,90,180){3}{6}
\Text(-43,40)[l]{$W$}
\Photon(0,40)(0,80){2}{4}
\end{picture}
\caption[]{\sf 
One-loop diagrams that give rise to neutrino magnetic moment in
standard model aided with right-handed neutrinos.  The lines marked
$\nu$ are generic neutrino lines, whereas those marked $\ell$ are
generic charged leptons.  The external vector boson line is the
photon.  In renormalizable gauges, there are extra diagrams where any
of the $W$ lines can be replaced by the corresponding unphysical Higgs
scalar. 
\label{f:magmom}}
\end{center}
\end{figure}
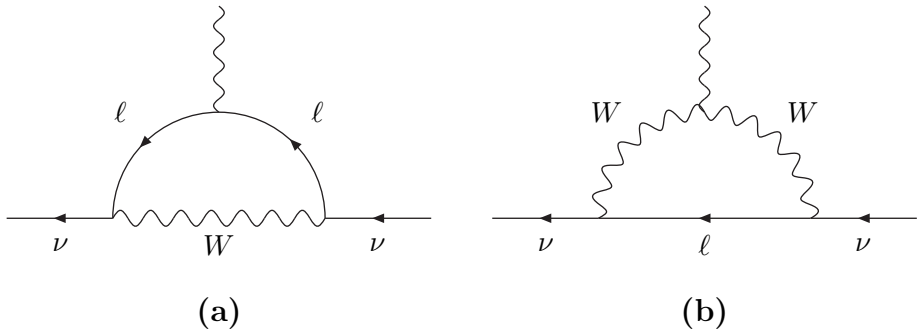

We now know that neutrinos are not massless as the standard model
presupposes.  Inclusion of neutrino mass naturally takes us beyond the
standard model, where the issue of neutrino interactions with a
magnetic field must be reassessed.  If the massive neutrino turns out
to be a Dirac fermion, its right-chiral projection must be included in
the fermion content of the theory, and in that case an anomalous
magnetic moment of a neutrino automatically emerges when quantum
corrections are taken into account.  In the simplest extension of the
standard model including right-chiral neutrinos, the magnetic moment
arises from the diagrams in Fig.~\ref{f:magmom} and is given
by~\cite{Fujikawa:yx}
\begin{eqnarray}
\mu_\nu = {3eG_F m_\nu \over 8\surd 2 \pi^2} = 3 \times 10^{-19} \mu_B
\times {m_\nu \overwithdelims() 1\,{\rm eV}} \,,
\label{munu}
\end{eqnarray}
where $m_\nu$ is the mass of the neutrino and $\mu_B$ is the Bohr
magneton.

%%%%%%%%%%%%%%%%
%%%%%%%%%%%%%%%%
If, on the other hand, neutrinos have Majorana masses,\footnote{For a
detailed discussion on Dirac and Majorana masses of neutrinos, see,
e.g., Ref.~\cite{Mohapatra:rq}.}  i.e., they are their own
antiparticles, they cannot have any magnetic moment at all, because
CPT symmetry implies that the magnetic moments of a particle and its
antiparticle should be equal and opposite.  However, even in this case
there can be transition magnetic moments, which are co-efficients of
effective operators of the form
$\bar\psi_1\sigma_{\mu\nu}\psi_2F^{\mu\nu}$, where $\psi_1$ and
$\psi_2$ denote two different fermion fields.  These will also
indicate some sort of interaction with the magnetic field, associated
with a change of the fermion flavor.

The question of the neutrino magnetic moment assumed immense
importance when it was suggested that it can be a potential solution
for the solar neutrino puzzle \cite{Cisneros:1970nq,Okun:na,Okun:hi}.
A viable solution required a neutrino magnetic moment around
$10^{-10}\mu_B$, orders of magnitude larger than that given by
\eqn{munu}, knowing that the neutrino masses cannot be very large.
However, such a magnitude could not be ruled out by direct laboratory
experiments.  A lot of research was carried out to explore possible
ways of evading the proportionality between the neutrino mass and
magnetic moment as shown in \eqn{munu}.  Voloshin
\cite{Voloshin:1987qy} showed that there may be symmetries to forbid
neutrino mass but not its magnetic moment.  Thus, if such a symmetry
remained unbroken, even massless neutrinos could have had a large
magnetic moment.  However, the proposed symmetries had to be broken in
the real world in order to meet other phenomenological constraints,
but in the end it was possible to envisage models where the ratio
between the magnetic moment and the mass of the neutrino is much
larger than that predicted by \eqn{munu}.\footnote{For an introduction
to such models, see e.g., Ref.~\cite{Pal:1991pm}.}

In this article, we will not follow the theoretical ideas outlined
above, mainly because the phenomenological motivation has become thin.
After the various recent solar neutrino experiments, especially the
data from the Sudbury Neutrino Observatory \cite{Ahmad:2002jz}, no one
believes that neutrino magnetic moments solve the solar neutrino
problem.  However, there may be celestial objects other than the sun
where the interaction between neutrinos and magnetic fields hold
the keys to some important questions.

It is not difficult to guess that the most important objects for this
purpose are the ones where very high magnetic fields are available.
Neutron stars have strong magnetic fields.  In fact, the surface
magnetic fields are typically of the order of $10^{12}$\,Gauss.  In
the core, the field might be larger.  Such high magnetic fields exist
also in the proto-neutron star, and its interaction with the neutrino
might have important effects on the supernova explosion.  There are
also objects called magnetars whose magnetic field is much higher than
that in ordinary neutron stars.  Even a small magnetic moment can have
a large effect in such systems.

But effects need not come through magnetic moment alone.  There may be
other physical quantities which, like the neutrino coupling to the
photon in Fig.~\ref{f:magmom}, contain charged particles in virtual
lines.  Calculation of such a diagram would be affected by a
background magnetic field through the propagator of virtual charged
particles, even if the external lines contain only neutrinos and
possibly other uncharged particles like the photon.  The simplest
example of physical quantities of this sort is the neutrino
self-energy.  Due to electrons in the internal lines, it is affected
by a background magnetic field.  Many other such examples can be
given, and some will be discussed later in this review.

We can also think of a different class of effects, where a process
involving neutrinos contains charged particles in either the initial
or the final state.  Since the asymptotic states of a charged particle
are affected by the presence of a magnetic field, the rates of such
processes would depend on the magnetic field.  This also can have
important physical implications.

On top of all these considerations, there is another very important
one.  If the background magnetic field is seeded in a material medium,
there can be extra effects coming from the density or the temperature
of the medium.  This also opens up many new interesting possibilities,
as we will see later in this review.

No matter which class of problems one considers, at a basic level one
must tackle the interaction of charged particles with magnetic
fields.  We therefore start with a short introduction to a field
theoretical discussion of charged particles in magnetic fields.

%%%%%%%%%%%%%%%%%%%%
\section{Charged particles in magnetic fields}
%%%%%%%%%%%%%%%%%%%%
\subsection{Spinor solutions}
%%%%%%%%%%%%%%%%%%%%
Unless otherwise mentioned, we will always talk about a homogeneous
and static magnetic field.  The background field tensor will be
denoted by $B_{\mu\nu}$, the magnetic field 3-vector by $\vec B$, and
its magnitude by $B$.  Without loss of generality, the magnetic field
can be assumed to direct in the $z$-direction.  In quantum theory, the
vector potential $\vec A$ would appear directly in the equations.  It
can be chosen in many equivalent ways.  For example, one can
choose\footnote{We employ the notation that a lettered subscript would
mean the contravariant component of a 4-vector.  If the covariant
component has to be used, it will be denoted by a numbered subscript.}
\begin{eqnarray}
A_0 = A_y = A_z = 0 \,, \qquad A_x = -yB\,,
\label{A}
\end{eqnarray}
or
\begin{eqnarray}
A_0 = A_x = A_z = 0 \,, \qquad A_y = xB\,,
\label{Aalt}
\end{eqnarray}
or more complicated ones where both $A_x$ and $A_y$ would be non-zero.
We will work with the choice of \eqn{A}.  The stationary state
solutions of the Dirac equations have the energy eigenvalues~\cite{JL49}
\begin{eqnarray}
E^2 = m^2 + p_z^2 + 2NeB \,,
\label{E}
\end{eqnarray}
where $N$ is a non-negative integer and $e$ is the positive unit of
charge, taken as usual to be equal to the proton charge.  For a fixed
value of $p_z$, the energy eigenvalues are thus quantized.  The
quantum number $N$ is called the Landau level, because \eqn{E} is the
generalization of a similar formula obtained by Landau in the
non-relativistic regime.  As is obvious from the energy relation, the
validity of the non-relativistic approximation requires not only that
the momentum must be small compared to the mass, but also $|eB|\ll
m^2$.  Since the lightest charged particle is the electron, the ratio
\begin{eqnarray}
B_e = m_e^2 / e = 4.4 \times 10^{13} \gauss
\label{Bc}
\end{eqnarray}
can be taken as a benchmark value for the magnetic field beyond which
relativistic effects cannot be ignored.\footnote{Many authors
denote this value by $B_c$ and call it the `critical field'.  This is
misleading.  There is nothing critical about this value.  Magnetic
field effects exist both below and above this value.}  Since the
potential applications involve stellar objects where the magnetic
fields can be comparable to, or larger than, this benchmark value
$B_e$, we will always use the relativistic formulas.

Note that the components of the momentum perpendicular to the magnetic
field do not enter the dispersion relation.  The eigenfunctions
corresponding to the positive and negative roots of $E$ can be written
as
\begin{eqnarray}
e^{-ip\cdot X {\omit y}} U_s (y,N,\vec p \omit y) \qquad \mbox{and}
\qquad 
e^{ip\cdot X {\omit y}} V_s (y,N,\vec p \omit y) \,, 
\label{UV}
\end{eqnarray}
where $U$ and $V$ are spinors, whose explicit forms will be given
shortly.  The coordinate 4-vector has been represented by $X^\mu$ (in
order to distinguish it from $x$, which is one of the components of
$X^\mu$).  The symbol $X^\mu\omit y$ stands for the same 4-vector,
with the difference that the $y$-coordinate has been set to zero.
Thus, for example,
\begin{eqnarray}
p\cdot X {\omit y} = Et - p_xx - p_zz \,.
%\label{}
\end{eqnarray}

The exponential factors in the eigenfunctions therefore do not contain
the $y$-coordinate.  However, the $y$-coordinate appears in the
spinor, along with the non-$y$ components of momentum.  For the
electron field, the components of the spinors can be conveniently
expressed in terms of the dimensionless variable $\xi$ defined by
\begin{eqnarray}
\xi = \sqrt{eB} \left( y - {p_x \over eB} \right) \,,
\label{xi}
\end{eqnarray}
and by defining the following function of $\xi$:
\begin{eqnarray}
I_N(\xi) = \left( {\sqrt{eB} \over N! \, 2^N \sqrt{\pi}} \right)^{1/2}
\, e^{-\xi^2/2} H_N(\xi) \,,
%\label{}
\end{eqnarray}
where $H_N(\xi)$ denote Hermite polynomials, and the normalizing
factor ensures that the functions $I_N(\xi)$ satisfy the following
completeness relation:
\begin{eqnarray}
\sum_N I_N(\xi) I_N(\xi_\star) = \sqrt{eB} \; \delta(\xi-\xi_\star)
= \delta (y-y_\star) \,.
\label{completeness}
\end{eqnarray}
In terms of these notations, it is now easy to write down the spinors
appearing in \eqn{UV}.  Using the shorthand
\begin{eqnarray}
M_N = \sqrt{2NeB} \,,
\end{eqnarray}
the $U$-spinors can be written as
\begin{eqnarray}
U_+ (y,N,\vec p \omit y) = \left( \begin{array}{c} 
I_{N-1}(\xi) \\[2ex] 0 \\[2ex] 
{\textstyle p_z \over \textstyle E_N+m} I_{N-1}(\xi) \\[2ex]
- {\textstyle M_N \over \textstyle 
E_N+m} I_N     (\xi) 
\end{array} \right) \,, \qquad 
U_- (y,N,\vec p \omit y) = \left( \begin{array}{c} 
0 \\[2ex] I_N (\xi) \\[2ex]
- {\textstyle M_N \over \textstyle E_N+m} I_{N-1}(\xi) \\[2ex]
-\,{\textstyle p_z \over \textstyle 
E_N+m} I_N(\xi) 
\end{array} \right) \,.
\label{Usoln}
\end{eqnarray}
While using this and other formulas for $N=0$, one should put
$I_{-1}=0$.  This implies that only the $U_-$ solution exists for
$N=0$.  Similarly, the $V$-spinors are given by
\begin{eqnarray}
V_+ (y,N,\vec p\omit y) = \left( \begin{array}{c} 
{\textstyle p_z \over \textstyle E_N+m} I_{N-1}(\widetilde\xi) \\[2ex]
{\textstyle M_N \over \textstyle E_N+m} I_N (\widetilde\xi)
\\[2ex] 
I_{N-1}(\widetilde\xi) \\[2ex] 0
\end{array} \right) \,, \qquad 
V_- (y,N,\vec p\omit y) = \left( \begin{array}{c} 
{\textstyle M_N \over \textstyle E_N+m} I_{N-1}(\widetilde\xi) \\[2ex]
-\,{\textstyle p_z \over \textstyle E_N+m} I_N(\widetilde\xi)
\\[2ex] 
0 \\[2ex] I_N (\widetilde\xi)
\end{array} \right) \,.
\label{Vsoln}
\end{eqnarray}
where
\begin{eqnarray}
\widetilde \xi=\sqrt{eB}\left(y+{p_x\over eB}\right) \,.
\end{eqnarray}
%

%%%%%%%%%%%%%%%%%%%%
\subsection{Propagator}
%%%%%%%%%%%%%%%%%%%%
In a field theoretic calculations, the spinors given above should be
used if the charged particle appears in the initial or the final state
of a physical process.   If, on the other hand, the charged particle
appears in the internal lines, we should use its propagator.  

There are two ways to write the propagator.  The first is to start
with the fermion field operator $\psi(X)$ written in terms of the
spinor solutions and the creation and annihilation operators, and
construct the time ordered product, as is usually done for finding the
propagator of a free fermion field in the vacuum.  The algebra is
straight forward and yields the result
\begin{eqnarray}
i S_B (X,X') = i \sum_N \int \frac{dp_0\, dp_x\, dp_z}{(2\pi)^3} \;
{E+m \over p_0^2 - E^2 + 
i\epsilon} e^{-ip \cdot (X\omit y - X'\omit y)} \nonumber\\*
\times \sum_s U_s (y,N,\vec p\omit y) \overline U_s (y', N,\vec p\omit
y) \,,
\label{Furryprop}
\end{eqnarray}
where $E$ is the positive root obtained from \eqn{E}.  The spin sum
can be conveniently written by introducing the following notation.
Given any vector $a^\mu$, we will define the following 4-vectors whose
components are given by
\begin{eqnarray}
a_\parallel^\mu &=& \big( a_0,0,0,a_z \big) 
\nonumber \\*
\widetilde a_\parallel^\mu &=& \big( a_z,0,0,a_0 \big) 
\nonumber \\*
a_\perp^\mu &=& \big( 0,a_x,a_y,0 \big) 
\end{eqnarray}
in the frame in which the background field is purely magnetic.  
Then, for any two 4-vectors $a$ and $b$, we will write
\begin{eqnarray}
a \cdot b_\parallel &=& a_\alpha b_\parallel^\alpha \,, 
\nonumber\\ 
a \cdot b_\perp &=& a_\alpha b_\perp^\alpha \,.
\label{parperp}
\end{eqnarray}
In this notation, the spin sum appearing in \eqn{Furryprop} can be
written as
\begin{eqnarray}
\sum_s U_s (y,N,\vec p\omit y) \overline U_s (y', N,\vec p\omit
y) = 
{1\over 2(E_N+m)} &\times& 
\bigg[ \left\{ m(1+\sigma_z) +
\rlap/p_\parallel - 
\widetilde{\rlap/p}_\parallel \gamma_5 \right\} I_{N-1}(\xi)
I_{N-1} (\xi') \nonumber\\*
&& + \left\{ m(1-\sigma_z) + \rlap/p_\parallel +
\widetilde{\rlap/p}_\parallel \gamma_5 \right\} I_N(\xi)
I_N(\xi') \nonumber\\*
&& - M_N (\gamma_1 - i\gamma_2) I_N(\xi) I_{N-1}(\xi') \nonumber\\*
&& - M_N (\gamma_1 + i\gamma_2) I_{N-1}(\xi) I_N(\xi') \bigg] \,,
\label{PU}
\end{eqnarray}
where
\begin{eqnarray}
\sigma_z \equiv i\gamma_1 \gamma_2 = - \gamma_0 \gamma_3 \gamma_5 \,.
\label{sigz} 
\end{eqnarray}
The resulting propagator is called the propagator in the Furry
picture.

Alternatively, one uses a functional procedure introduced by Schwinger
\cite{Schwinger:nm} where the propagator is written in the form
\begin{eqnarray}
iS_B(X,X') = \Psi(X,X') \int \frac{d^4 p}{(2 \pi)^4} e^{-ip \cdot
(X-X')} iS_B (p) \,,
\label{schwingprop}
\end{eqnarray}
where $S_B(p)$ is expressed as an integral over a variable $s$,
usually (though confusingly) called the `proper time':
\begin{eqnarray}
i S_B (p) =\int_0^\infty ds\; e^{\Phi(p,s)} G(p,s) \,.
\label{SB}
\end{eqnarray}
The quantities $\Phi(p,s)$ and $G(p,s)$ can be written in the
following way, using the notation of \eqn{parperp}:
\begin{eqnarray}
\Phi(p,s) &\equiv& is \left( p_\parallel^2 - {\tan
eBs \over eBs} \, p_\perp^2 - m^2 \right) - \epsilon
|s| \,, 
\label{Phi}
\\
G(p,s) &\equiv&  {e^{ieBs\sigma\!_z} \over \cos eBs} \;
\left( 
\rlap/p_\parallel - {e^{-ieBs\sigma_z} \over \cos eBs}
\rlap/ p_\perp + m \right) \nonumber\\*
&=& ( 1 + i\sigma_z \tan  eBs ) 
(\rlap/p_\parallel + m ) - (\sec^2 eBs) \rlap/ p_\perp \,,
\label{G}
\end{eqnarray}
In a typical loop diagram, one therefore will have to perform not only
integrations over the loop momenta, but also over the proper time
variables.

The other factor $\Psi(X,X')$ appearing in Eq.\,(\ref{schwingprop}) is
a phase factor which breaks translation invariance and is given
by~\cite{Schwinger:nm} 
\begin{eqnarray}
\Psi(X,X') = \exp\left( ie \int_{X'}^X d\xi^\mu \left[ A_\mu(\xi) 
+ \frac12 B_{\mu\nu} (\xi^\nu - X'^\nu) \right] \right) \,.
\label{Psi}
\end{eqnarray}
The integral is path-independent.  The second term does not contribute
if one chooses a straight line path characterized by
\begin{eqnarray}
\xi^\mu = (1-\lambda) X'^\mu + \lambda X^\mu \,, \qquad 0 \leq \lambda
\leq 1 \,.
%\label{}
\end{eqnarray}
Further, the vector potential for a constant field $B_{\mu\nu}$ can be
written as
\begin{eqnarray}
A_\mu (\xi) = - \frac12 B_{\mu\nu} \xi^\nu \,.
%\label{}
\end{eqnarray}
The integration in \eqn{Psi} can then be performed easily and one
obtains 
\begin{eqnarray}
\Psi(X,X') = \exp \left(- \frac12 ie X^\mu B_{\mu\nu} X'^\nu \right) \,.
\label{xBx}
\end{eqnarray}
In what follows, we will indicate where this phase factor cancels
between different propagators, and where it does not.

It is not difficult to obtain the modification of the propagator if
the charged particle is in a background magnetized plasma.  In this
case, the background contains both matter and magnetic field.  The
clue can now be obtained from propagator in thermal matter without any
magnetic field.  In the real-time formalism, the propagator $iS'(p)$
involving the time-ordered product\footnote{It should be mentioned
here that other orderings also appear in the evaluation of general
Green's functions.  We will not talk about these other propagators.}
can be written in terms of the free propagator $iS_0(p)$:
\begin{eqnarray}
iS'(p) = iS_0(p) - \eta_F(p) \Big[ iS_0(p) - i\overline S_0(p) \Big] \,,
%\label{}
\end{eqnarray}
where 
\begin{eqnarray}
\overline S_0(p) = \gamma_0 S_0^\dagger(p) \gamma_0 \,,
%\label{}
\end{eqnarray}
and $\eta_F(p)$ contains the distribution function for particles and
antiparticles:
\begin{eqnarray}
\eta_F(p) = \Theta(p\cdot u) f_F(p,\mu,\beta) 
+ \Theta(-p\cdot u) f_F(-p,-\mu,\beta) \,.
\label{eta}
\end{eqnarray}
Here, $\Theta$ is the step function which takes the value $+1$ for
positive values of its argument and vanishes for negative values of
the argument, $u^\mu$ is the 4-vector denoting the center-of-mass
velocity of the background plasma, and $f_F$ denotes the Fermi-Dirac
distribution function:
\begin{eqnarray}
f_F(p,\mu,\beta) = {1\over e^{\beta(p\cdot u - \mu)} + 1} \,.
\end{eqnarray}

In a similar manner, the propagator in a magnetized plasma is given
by~\cite{Elmfors:1996gy}
\begin{eqnarray}
iS'_B(p) = iS_B(p) - \eta_F(p) \Big[ iS_B(p) - i\overline S_B(p)
\Big] \,.
%\label{fullprop}
\end{eqnarray}
In the Schwinger proper-time representation, this can also be written
as an integral over the proper-time variable $s$:
\begin{eqnarray}
iS'_B(p) = \int_0^\infty ds\; e^{\Phi(p,s)} G(p,s) 
- \eta_F(p) \int_{-\infty}^\infty ds\; 
e^{\Phi(p,s)} G(p,s) \,,
\label{fullprop}
\end{eqnarray}
where $\Phi(p,s)$ and $G(p,s)$ are given by the expressions in
\eqn{Phi} and \eqn{G}.

%%%%%%%%%%%%%%%%
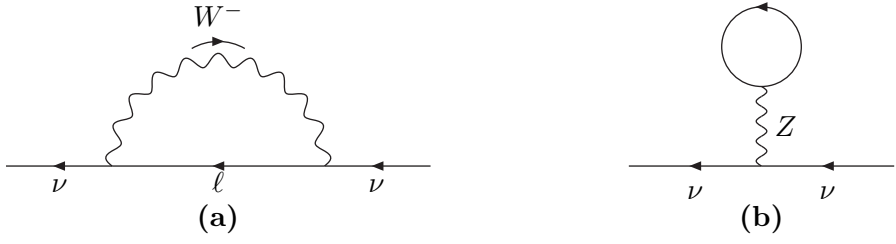
\begin{figure}[btp]
\begin{center}
\begin{picture}(180,75)(-90,-20)
\ArrowLine(80,0)(40,0) 
\Text(60,-10)[b]{$\nu$} 
\ArrowLine(40,0)(-40,0)
\Text(0,-10)[b]{$\ell$} 
\ArrowLine(-40,0)(-80,0)
\Text(-60,-10)[b]{$\nu$} 
\PhotonArc(0,0)(40,0,180){2.5}{10.5}
\ArrowArcn(0,27)(20,120,60)
\Text(0,53)[b]{$W^-$}
\Text(0,-20)[]{\bf (a)}
\end{picture}
\qquad
\begin{picture}(180,75)(-90,-20)
\ArrowLine(50,0)(0,0) 
\Text(25,-10)[c]{$\nu$} 
\ArrowLine(0,0)(-50,0)
\Text(-25,-10)[c]{$\nu$} 
\Photon(0,0)(0,30)24
\Text(5,15)[l]{$Z$}
\ArrowArc(0,45)(15,-90,270)
\Text(0,-20)[]{\bf (b)}
\end{picture}
\caption[]{\sf One-loop diagrams for neutrino self-energy in a
magnetized medium.  Diagram b is absent if the background contains
only a magnetic field but no matter.  For legends and related
diagrams, see the caption of Fig.~\ref{f:magmom}.
\label{f:selfen}}
\end{center}
\end{figure}
%%%%%%%%%%%%%%%%
%%%%%%%%%%%%%%%%%%%%
\section{Magnetic field effects on neutrinos from virtual charged
particles}\label{vi}
%%%%%%%%%%%%%%%%%%%%
\subsection{Neutrino self-energy}\label{vise}
%%%%%%%%%%%%%%%%%%%%
We have already mentioned in Sec.~\ref{mo} that the simplest physical
quantity where background magnetic field effects appear through
virtual lines of charged particles is the self energy of the neutrino.
The 1-loop diagram for the self energy is given in
Fig.~\ref{f:selfen}.

It is easy to see how the self-energy might be modified within a
magnetized plasma.  In the vacuum, the self-energy of a fermion has
the general structure
\begin{eqnarray}
\Sigma(p) = a \gamma^\mu k_\mu + b \,,
%\label{}
\end{eqnarray}
which is the most general form dictated by Lorentz covariance.  Here,
$a$ and $b$ are Lorentz invariant, and can therefore depend only on
$k^2$.  In the presence of a homogeneous medium, the self-energy will
involve the 4-vector $u^\mu$ introduced in \eqn{eta}.  Further, if
the medium contains a background magnetic field, the background field
$B_{\mu\nu}$ also enters the general expression for the self-energy.
These new objects, $u^\mu$ and $B_{\mu\nu}$, enter in two different
ways.  First, any form factor now can depend on more Lorentz
invariants which are present in the problem.  Second, the number of
form factors also increases, since it is possible to write some more
Lorentz covariant terms using $u^\mu$ and $B_{\mu\nu}$.  There will in
fact be a lot of form factors in the most general case.  However, if
we have chiral neutrinos as in the standard electroweak theory, the
expression is not very complicated:
\begin{eqnarray}
\Sigma_B(p) = 
\Big( a_1 k_\mu + b_1 u_\mu 
+ a_2 k^\nu B_{\mu\nu} + b_2 u^\nu B_{\mu\nu} 
+ a_3 k^\nu \widetilde B_{\mu\nu} + b_3 u^\nu \widetilde B_{\mu\nu} 
\Big) \gamma^\mu L \,, 
\label{Sigma}
\end{eqnarray}
where $L$ is the left-chiral projection operator, and
\begin{eqnarray}
\widetilde B_{\mu\nu} = \frac12 \epsilon_{\mu\nu\lambda\rho}
B^{\lambda\rho} \,. 
%\label{}
\end{eqnarray}

We first consider the self-energy when the background consists of a
pure magnetic field, without any matter.  Then all $b$-type
form-factors disappear from the self-energy.  The dispersion relation
of neutrinos can then be obtained by the zeros of $\rlap/k-\Sigma_B$,
which gives
\begin{eqnarray}
\Big[ (1-a_1) k^\mu - a_2 k^\nu B_{\mu\nu} 
- a_3 k^\nu \widetilde B_{\mu\nu} \Big]^2 = 0 \,.
%\label{}
\end{eqnarray}
Performing the square is trivial, and one obtains
\begin{eqnarray}
(1-a_1)^2 k^\mu k_\mu 
+ a_2^2 k^\nu k_\lambda B_{\mu\nu} B^{\mu\lambda} 
+ a_3^2 k^\nu k_\lambda \widetilde B_{\mu\nu} \widetilde B^{\mu\lambda} 
+ 2a_2 a_3 k^\nu k_\lambda B_{\mu\nu} \widetilde B^{\mu\lambda} =0 \,.
%\label{}
\end{eqnarray}
It is interesting to note that the terms linear in the background
field all vanish due to the antisymmetry of the field tensor.
Moreover, the $a_2a_3$ term is also zero for a purely magnetic field.

The remaining terms can be most easily understood if we take the
$z$-axis along the direction of the magnetic field.  Then the
only non-zero components of the tensor $B_{\mu\nu}$ and $\widetilde
B_{\mu\nu}$ are given by
\begin{eqnarray}
B_{12} = - B_{21} = B \,, \qquad
\widetilde B_{03} = - \widetilde B_{30} = B \,,
%\label{}
\end{eqnarray}
where we have adopted the convention
\begin{eqnarray}
\epsilon_{0123} = +1 \,.
%\label{}
\end{eqnarray}
Thus
\begin{eqnarray}
k^\nu k_\lambda B_{\mu\nu} B^{\mu\lambda} &=& - (k_x^2 + k_y^2) B^2
= - k_\perp^2 B^2 \,,
\nonumber\\*
k^\nu k_\lambda \widetilde B_{\mu\nu} \widetilde B^{\mu\lambda} &=& 
(\omega^2 - k_z^2) B^2 = k_\parallel^2 B^2 \,,
%\label{Sigma_B}
\end{eqnarray}
where the notations for parallel and perpendicular products were
introduced in \eqn{parperp}.  The form factor $a_1$ can be set equal
to zero by a choice of the renormalization prescription.  So the
dispersion relation is now a solution of the equation
\begin{eqnarray}
k^2 - a_2^2 k_\perp^2 B^2 + a_3^2 k_\parallel^2 B^2 =0 \,,
%\label{}
\end{eqnarray}
which can also be written as
\begin{eqnarray}
\omega^2 = k_z^2 + {1 + a_2^2 B^2 \over 1 + a_3^2 B^2} \, k_\perp^2 \,.
%\label{}
\end{eqnarray}
Of course, this should not be taken as the solution for the neutrino
energy, because the right hand side contains form factors which, in
general, are functions of the energy and other things.  But at least
it shows that in the limit $B\to0$, the vacuum dispersion relation is
recovered.  If we retain the lowest order corrections in $B$, we can
treat the form-factors to be independent of $B$ and write
\begin{eqnarray}
\omega^2 = \vec k^2 + (a_2^2 - a_3^2) B^2 k_\perp^2 \,.
\label{disp_B}
\end{eqnarray}
Calculation of this self-energy was performed by Erdas and Feldman
\cite{Erdas:gy} using the Schwinger propagator, where they also
incorporated the modification of the $W$-propagator due to the
magnetic field.  Importantly, the $W$-propagator contains the same
phase factor as given in \eqn{xBx}.  Therefore, the phase
factors from the charged lepton and the $W$-lines are of the form
$\Psi(X,X')\Psi(X',X)$.  From \eqn{xBx}, it is easy to see that this
is equal to unity, and therefore the phase factors do not contribute
in the final expression.  Detailed calculations show
that~\cite{Erdas:gy}
\begin{eqnarray}
a_2^2 - a_3^2 = {eg \overwithdelims() 2\pi M_W^2}^2
\left( \frac13 \ln {M_W \over m} + \frac18 \right) \,,
\label{a2sq-a3sq}
\end{eqnarray}
where $m$ is the mass of the charged lepton in the internal line.  For
strong magnetic fields, the dispersion relation has been calculated
more recently by Elizalde, Ferrer and de la
Incera~\cite{Elizalde:2000vz}.

Let us next concentrate on the terms which can occur only in a
magnetized medium.  In other words, we select out the terms which
cannot occur if the neutrino propagates in a background of pure
magnetic field without any material medium.  This means that, apart
from the term $\rlap/k$ which occurs also in the vacuum, we look for
the terms which contain both $u^\mu$ and $B_{\mu\nu}$.  Further, if
the background field is purely magnetic in the rest frame of the
medium, $u^\nu B_{\mu\nu}=0$ since $u$ has only the time component
whereas the only non-zero components of $B_{\mu\nu}$ are spatial.
Thus we are left with~\cite{D'Olivo:1989cr}:
\begin{eqnarray}
\Sigma_B(p) = 
\Big( a_1 k_\mu + b_1 u_\mu + b_3 u^\nu \widetilde B_{\mu\nu} 
\Big) \gamma^\mu L \,.
\label{Sigma_uB}
\end{eqnarray}
Once again, setting $a_1=0$ through a renormalization prescription, we
can find the dispersion relation of the neutrinos in the
form~\cite{D'Olivo:1989cr}: 
\begin{eqnarray}
\omega = \Big| \vec k - b_3 \vec B \Big| + b_1 \approx |\vec k| -
b_3 \hat k \cdot \vec B + b_1 \,,
\label{magdisp}
\end{eqnarray}
where $\hat k$ is the unit vector along $\vec k$, and we have kept
only the linear correction in the magnetic field.  This form for the
dispersion relation was first arrived at by D'Olivo, Nieves and Pal
(DNP) \cite{D'Olivo:1989cr} who essentially performed a calculation to
the first order in the external field.  As for the form factors, $b_1$
was known previously, obtained from the analysis of neutrino
propagation in isotropic matter, i.e., without any magnetic field.
The result was~\cite{Wolfenstein:1977ue, Notzold:1987ik, Pal:1989xs,
Nieves:ez}
\begin{eqnarray}
b_1 &=& \surd2 G_F (n_e - n_{\bar e}) \times \left(y_e +
\rho c_V \right) \,,  
\label{b1}
\end{eqnarray}
where 
\begin{eqnarray}
\rho = {M_W^2 \over M_Z^2\cos^2\theta_W} \,,
%\label{}
\end{eqnarray}
$n_e$, $n_{\bar e}$ are the densities of electrons and positrons
in the medium, 
\begin{eqnarray}
y_e = \cases{1 & for $\nu_e$, \cr
0 & for $\nu\neq\nu_e$,}
%\label{}
\end{eqnarray}
and $c_V$ is defined through the coupling of the electron to the
$Z$-boson, whose Feynman rule is
\begin{eqnarray}
-{ig \over 2\cos\theta_W} \; \gamma_\mu (c_V - c_A\gamma_5) \,.
%\label{}
\end{eqnarray}
In other words, in the standard model
\begin{eqnarray}
c_V = - \frac12 + 2 \sin^2 \theta_W \,, \qquad c_A = - \frac12 \,.
\label{cvca}
\end{eqnarray}

The contribution to $b_3$ from background electrons and positrons was
calculated by DNP \cite{D'Olivo:1989cr}.  They obtained\footnote{The
authors of Ref.~\cite{D'Olivo:1989cr} used a convention in which
$e<0$.  Here we present the result in the convention $e>0$.}
\begin{eqnarray}
b_3 &=& -2\surd2eG_F 
\int {d^3p \over (2\pi)^3 2E} \; {d\over dE} (f_e - f_{\bar e})
\times \left(y_e + \rho c_A \right) \,, 
\label{b3}
\end{eqnarray}
where $f_e$ and $f_{\bar e}$ are the Fermi distribution functions for
electrons and positrons, and
\begin{eqnarray}
E = \sqrt{\vec p^2 + m_e^2} \,.
%\label{}
\end{eqnarray}
Later authors have improved on this result in two different ways.
Some authors \cite{D'Olivo:1997vi} have included the contributions
coming from nucleons in the background.  Some others 
\cite{Elmfors:1996gy,Erdas:1998uu} have used the
Schwinger propagator and extended the results to all orders in the
magnetic field.

%%%%%%%%%%%%%%%%%%%%
\subsection{Neutrino mixing and oscillation}\label{vios}
%%%%%%%%%%%%%%%%%%%%
Calculation of neutrino self-energy has a direct consequence on
neutrino mixing and oscillations.  Of course neutrino oscillations
require neutrino mixing and therefore neutrino mass.  For the sake of
simplicity, we discuss mixing between two neutrinos which we will call
$\nu_e$ and $\nu_\mu$.  The eigenstates will in general be called
$\nu_1$ and $\nu_2$, which are given by
\begin{eqnarray}
{\nu_1 \choose \nu_2} = \left( \begin{array}{ccc}
\cos\theta && -\sin\theta \\
\sin\theta && \cos\theta
			\end{array} \right)
{\nu_e \choose \nu_\mu} \,.
%\label{}
\end{eqnarray}
We will denote the masses of the eigenstates by $m_1$ and $m_2$, and
assume that the neutrinos are ultra-relativistic.  Then in the vacuum,
the evolution equation for a beam of neutrinos will be given by
\begin{eqnarray}
i{d\over dt} {\nu_e \choose \nu_\mu} = {1\over 2\omega} M^2 
{\nu_e \choose \nu_\mu} \,.
%\label{}
\end{eqnarray}
where the matrix $M^2$ is given by
\begin{eqnarray}
M^2 = \left( \begin{array}{ccc}
-\frac12 \Delta m^2 \cos 2\theta && \frac12 \Delta m^2 \sin 2\theta \\
\frac12 \Delta m^2 \sin 2\theta && \frac12 \Delta m^2 \cos 2\theta
			\end{array} \right) \,,
%\label{}
\end{eqnarray}
where $\Delta m^2=m_2^2-m_1^2$.  In writing this matrix, we have
ignored all terms which are multiples of the unit matrix, which affect
the propagation only by a phase which is common for all the states.

In a non-trivial background, the dispersion relations of the neutrinos
change, as discussed in Sec.~\ref{vise}.  This adds new terms to the
diagonal elements of the effective Hamiltonian in the flavor basis,
which we denote by the symbol $A$.  As a result, the matrix $M^2$
should now be replaced by
\begin{eqnarray}
\widetilde M^2 = \left( \begin{array}{ccc}
-\frac12 \Delta m^2 \cos 2\theta + A_{\nu_e} 
&& \frac12 \Delta m^2 \sin 2\theta \\
\frac12 \Delta m^2 \sin 2\theta 
&& \frac12 \Delta m^2 \cos 2\theta + A_{\nu_\mu}
			\end{array} \right) \,,
\label{MM}
\end{eqnarray}
where the extra contributions are in general different for $\nu_e$ and
$\nu_\mu$.  The eigenstates and eigenvalues change because of these
new contribution.  For example, the mixing angle now becomes
$\widetilde\theta$, given by
\begin{eqnarray}
\tan 2\widetilde\theta = {\Delta m^2 \sin 2\theta \over \Delta m^2
\cos 2\theta + A_{\nu_\mu} - A_{\nu_e} } \,.
\label{RC}
\end{eqnarray}

In a pure magnetic field, the self-energies were shown in
\eqn{disp_B} and \eqn{a2sq-a3sq}.  The quantity $m$ appearing in
\eqn{a2sq-a3sq} is the mass of the charged lepton in the loop.  Thus,
for $\nu_e$, it is the electron mass whereas for $\nu_\mu$, it is the
muon mass.  Thus $A_{\nu_\mu} \neq A_{\nu_e}$.   However, the
difference appears in logarithmic form, and is presumably not very
significant.

In a magnetized medium, however, the situation changes.  The reason is
that the medium contains electrons but not muons.  Accordingly, the
quantities $A_{\nu_e}$ and $A_{\nu_\mu}$ can be very different, as
seen by the presence of the term $y_e$ in \eqn{b3}.  If we take
self-energy corrections only up to linear order in $B$, as done in
\eqn{magdisp}, we obtain
\begin{eqnarray}
A_{\nu_\mu} - A_{\nu_e} = -\surd2 G_F (n_e - n_{\bar e}) - 2\surd2
eG_F \hat k \cdot \vec B 
\int {d^3p \over (2\pi)^3 2E} \; {d\over dE} (f_e - f_{\bar e})
\,.
\label{A-A}
\end{eqnarray}
The first term on the right side comes just from the background
density of matter, and the second term is the magnetic field dependent
correction.  This quantity has been calculated for various
combinations of temperature and chemical potential of the background
electrons \cite{Elmfors:1996gy, Esposito:1995db, D'Olivo:1995bq}.

If the denominator of the right side of \eqn{RC} becomes zero for some
value of $A_{\nu_\mu} - A_{\nu_e}$, the value of
$\tan2\widetilde\theta$ will become infinite.  This is the resonant
level crossing condition.  This was first discussed in the context of
neutrino oscillation in a matter background by Mikheev and Smirnov
\cite{Mikheev:wj}, where a particular value of density would ensure
resonance.  Presence of a magnetic field will modify this resonant
density, as seen from \eqn{A-A}.  The modification will be direction
dependent because of the factor $\hat k \cdot \vec B$.  Some early
authors \cite{Esposito:1995db, D'Olivo:1995bq} contemplated that, for
large $B$, the magnetic term might even drive the resonance.  However,
later it was shown \cite{Nunokawa:1997dp} that the magnetic correction
would always be smaller than the other term.  So, if one considers
values of $B$ which are so large that the last term in \eqn{A-A} is
larger than the first term on the right hand side, it means that one
must take higher order corrections in $B$ into account.

It should be noted that the type of corrections to the dispersion
relation discussed in Sec.~\ref{vise} appear from chiral neutrinos.
Thus, they produce chirality-preserving modifications to neutrino
oscillations.  In addition, if the neutrino has a magnetic moment,
there will be chirality-flipping modifications as well.  Many of these
modifications were analyzed in the context of the solar neutrino
problem, and we do not discuss them here.\footnote{A recent paper on
chirality-flipping oscillations is Ref.~\cite{Egorov:1999ah}, where
one can obtain references to earlier literature.  Some early
references are also found in Refs.~\cite{Mohapatra:rq} and
\cite{Pal:1991pm}.}  As pointed out in Sec.~\ref{mo}, they are not
important for solar neutrinos, although may be important in other
stellar objects like the neutron star where the magnetic fields are
much larger.

%%%%%%%%%%%%%%%%%%%%
\subsection{Electromagnetic vertex of neutrinos}\label{viem}
\begin{figure}[t]
\begin{center}
\begin{picture}(100,100)(-50,-20)
\ArrowLine(50,0)(0,10) 
\Text(25,-2)[c]{$\nu$} 
\ArrowLine(0,10)(-50,0)
\Text(-25,-2)[c]{$\nu$} 
\ArrowArc(0,30)(20,-180,180)
\Text(22,40)[l]{$\ell$}
\Photon(0,50)(0,80)24
\Text(5,65)[l]{$\gamma$}
\end{picture}
\caption[]{\sf The neutrino electromagnetic vertex in the leading
order in the Fermi constant.
\label{f:4fermi}}
\end{center}
\end{figure}
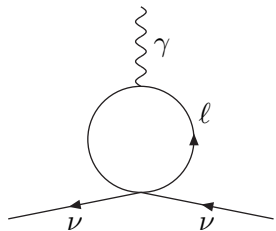

%%%%%%%%%%%%%%%%%%%%
A lot of work has also gone into evaluating the
neutrino-neutrino-photon vertex in the presence of a background
magnetic field.  The vertex arises from the diagrams of
Fig.~\ref{f:magmom}, which contain internal $W$-lines.  In addition,
there is a diagram mediated by the $Z$-boson, as shown in
Fig.~\ref{f:Zvertex}.  For phenomenological purposes, we require the
electromagnetic vertex of neutrinos only in the leading order in Fermi
constant.  It should be realized that in this order, the diagram of
Fig.~\ref{f:magmom}b does not contribute at all, since it has two
$W$-propagators.  The remaining diagrams, shown in
Fig.~\ref{f:magmom}a and Fig.~\ref{f:Zvertex}, can both be represented
in the form shown in Fig.~\ref{f:4fermi}, where an effective 4-fermi
vertex has been used.  The effective 4-fermi interaction can be
written as
\begin{eqnarray}
\mathscr L_{\rm eff} = 
-\surd2 G_F \Big[ \bar\nu_L \gamma^\lambda \nu_L \Big]
\Big[ \bar\ell \gamma_\lambda (g_V - g_A\gamma_5) \ell \Big] \,.
\label{4fermi}
\end{eqnarray}
If the neutrino and the charged lepton belong to different generations
of fermions, this effective Lagrangian contains only the neutral
current interactions, and in that case $g_V$ and $g_A$ are identical
to $c_V$ and $c_A$ defined in \eqn{cvca}.  On the other hand, if both
$\nu$ and $\ell$ belong to the same generation, we should add the
charged current contribution as well, and use
\begin{eqnarray}
g_V = c_V +1  \,, \qquad g_A = c_A + 1 \,.
%\label{}
\end{eqnarray}
%

%%%%%%%%%%%%%%%%
\begin{figure}[t]
\begin{center}
\begin{picture}(180,80)(-90,-20)
\ArrowLine(50,0)(0,0)
\Text(25,-10)[c]{$\nu$} 
\ArrowLine(0,0)(-50,0)
\Text(-25,-10)[c]{$\nu$} 
\Photon(0,0)(0,20)23
\Text(5,10)[l]{$Z$}
\ArrowArc(0,30)(10,-180,180)
\Text(15,30)[l]{$\ell$}
\Photon(0,40)(0,60)23
\Text(5,65)[l]{$\gamma$}
\end{picture}
\caption[]{\sf $Z$-photon mixing diagram contributing to the neutrino
electromagnetic vertex.
\label{f:Zvertex}}
\end{center}
\end{figure}
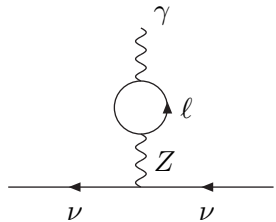
%%%%%%%%%%%%%%%%
%%%%%%%%%%%%%%%%
%%%%%%%%%%%%%%%%
Many processes involving neutrinos and photons have been calculated
using the 4-fermi Lagrangian of \eqn{4fermi}.  The calculations
simplify in this limit for various reasons.  First, we do not have to
use the momentum dependence of the gauge boson propagators.  Second,
since two charged lepton lines form a loop in Fig.~\ref{f:4fermi}, the
phase factor of \eqn{xBx} appearing in their propagators cancel each
other. 

The background magnetic field can give rise to many physical processes
which are impossible to occur in the vacuum.  One such process is the
decay of a photon into a neutrino-antineutrino pair:
\begin{eqnarray}
\gamma \to \nu + \bar\nu \,.
\label{gammadk}
\end{eqnarray}
This was calculated using the Schwinger propagator in some very early
papers \cite{GN72, DeRaad:1976kd}.  Assuming two generations of
fermions, the decay rate was found to be
\begin{eqnarray}
\Gamma = {\alpha^2 G_F^2 \over 48\pi^3 \Omega} \Big|
\varepsilon^\mu q^\nu \widetilde B_{\mu\nu} \Big|^2  \Big| \mathscr M_e-
\mathscr M_\mu \Big|^2 \,,
%\label{}
\end{eqnarray}
where $\varepsilon^\mu$, $q^\mu$ and $\Omega$ are the polarization
vector, the momentum 4-vector and the energy of the initial photon,
and the quantity $\mathscr M_\ell$ was evaluated in various limits by
these authors.  For example, if $\Omega\ll m_\ell$, they found
\begin{eqnarray}
\mathscr M_\ell = {\Omega^2 \over eB} \sin^2\theta \times 
\cases{
{2\over 15} \left({eB \over m_\ell^2} \right)^3 & for $eB \ll m_\ell^2$\,,
\cr  
{1\over 3} \left( {eB \over m_\ell^2} \right) & for $eB \gg m_\ell^2$\,,}
%\label{}
\end{eqnarray}
where $\theta$ is the angle between the photon momentum and the
magnetic field.  No matter which neutrino pair the photon decays to,
both charged leptons appear in the decay rate because of the loop in
Fig.~\ref{f:Zvertex}.  The calculation has been carried out in the
leading order in Fermi constant, where only the axial couplings of the
charged leptons contribute to the amplitude.

A related process is Cherenkov radiation from neutrinos:
\begin{eqnarray}
\nu \to \nu +\gamma \,.
\label{cheren}
\end{eqnarray}
Again, this is a process forbidden in the vacuum.  But a background
magnetic field modifies the photon dispersion relation, and so this
process becomes feasible.  The rate of this process has been
calculated by many authors \cite{GN72, Sko76, Ioannisian:1996pn,
Gvozdev:1997mc}, and all of them do not get the same result.  
According to Ref.~\cite{Gvozdev:1997mc}, the rate for the process is
given by
\begin{eqnarray}
\Gamma = {\alpha G_F^2\over 8\pi^2} (g_V^2 + g_A^2) (eB)^2
\omega\sin^2\theta F(\omega^2\sin^2\theta/eB) \,,
%\label{}
\end{eqnarray}
where $\omega$ is the initial neutrino energy and $\theta$ is the
angle between the initial neutrino momentum and the background
field.  For large magnetic fields satisfying the condition $eB \gg
\omega^2\sin^2\theta$, the function $F$ is given by
\begin{eqnarray}
F(x) = 1 - {x \over 2} + {x^2 \over 3} - {5x^3 \over 24} + {7x^4 \over
60} + \cdots \,.
%\label{}
\end{eqnarray}
The modification of this process in the presence of background matter
has also been calculated~\cite{Chistyakov:1999ii}.

Another process that has been discussed is the radiative neutrino
decay
\begin{eqnarray}
\nu_a \to \nu_b + \gamma \,.
%\label{}
\end{eqnarray}
Unlike the previous processes, this can occur in the vacuum as well
when the neutrinos have mass and mixing.  However, a background
magnetic field adds new contributions to the amplitude, and the rate
can be enhanced.  Gvozdev, Mikheev and Vasilevskaya
\cite{Gvozdev:1996kx} calculated the rate of this decay in a variety
of situations depending on the field strength and the energy of the
initial neutrino.  For a strong magnetic field ($B\gg B_e$), they
found the decay rate of an ultra-relativistic neutrino of energy
$\omega$ to be
\begin{eqnarray}
\Gamma = {2\alpha G_F^2 \over \pi^4} {m_e^6 \over \omega} {B
\overwithdelims() B_e}^2 |K_{ae}K_{be}^*|^2 J(\omega\sin\theta/2m_e) \,,
%\label{}
\end{eqnarray}
where $K$ is the leptonic mixing matrix, $\theta$ is the angle between
the magnetic field and the neutrino momentum, and
\begin{eqnarray}
J(z) = \int_0^z dy \; (z-y) \left( {1 \over y\sqrt{1-y^2}} \tan^{-1}
{y \over \sqrt{1-y^2}} - 1 \right)^2 \,.
%\label{}
\end{eqnarray}
The curious feature of this result is that this is independent of the
initial and the final neutrino masses.

The form for the 4-fermi interaction in \eqn{4fermi} suggests that the
neutrino electromagnetic vertex function $\Gamma_\lambda$ can be
written as~\cite{Nieves:1993er}:
\begin{eqnarray}
\Gamma_\lambda = - \, {\surd2 G_F \over e} \gamma^\rho L \Big( g_V
\Pi_{\lambda\rho} - g_A \Pi^5_{\lambda\rho} \Big) \,.
%\label{}
\end{eqnarray}
Here, the term $\Pi_{\lambda\rho}$ is exactly the expression for the
vacuum polarization of the photon, and appears from the vector
interaction in the effective Lagrangian.  The other term,
$\Pi^5_{\lambda\rho}$ differs from $\Pi_{\lambda\rho}$ in that it
contains an axial coupling from the effective Lagrangian.

This equality is valid even when one has a magnetic field and a
material medium as the background, as long as one restricts oneself to
the leading order in Fermi constant.  Thus, the calculation of the
photon self-energy in a background magnetic field in matter can give
us information about the neutrino electromagnetic vertex in the same
situation.  The calculation of the photon self-energy was done to the
first order in $B$ by Ganguly, Konar and Pal \cite{Ganguly:1999ts}.
Later, it was extended to all orders by D'Olivo, Nieves and Sahu
\cite{D'Olivo:2002sp}.  The calculation of $\Pi^5_{\lambda\rho}$, on
the other hand, was undertaken in a series of papers by Bhattacharya,
Ganguly, Konar and Das \cite{Bhattacharya:2001nm, Bhattacharya:2002ci,
Konar:2002gy}.  In particular, it was shown \cite{Bhattacharya:2002ci}
that the terms which are odd in $B$ contribute to the vertex function
even at zero momentum transfer, which means that they contribute to an
effective charge of the neutrino.

%%%%%%%%%%%%%%%%%%%%
\subsection{Neutrino-photon scattering}
%%%%%%%%%%%%%%%%%%%%
Gell-Mann showed \cite{Gell-Mann} that the amplitude of the reaction
\begin{eqnarray}
\gamma + \nu \to \gamma + \nu
\label{gnugnu}
\end{eqnarray}
is exactly zero to order $G_F$ because by Yang's
theorem \cite{Yang, Landau} two
photons cannot couple to a $J = 1$ state.  In the standard model,
therefore, amplitude of the above process appears only at the level of
$1/M_W^4$ and as a result the cross-section is exceedingly
small~\cite{Dicus:iy}.

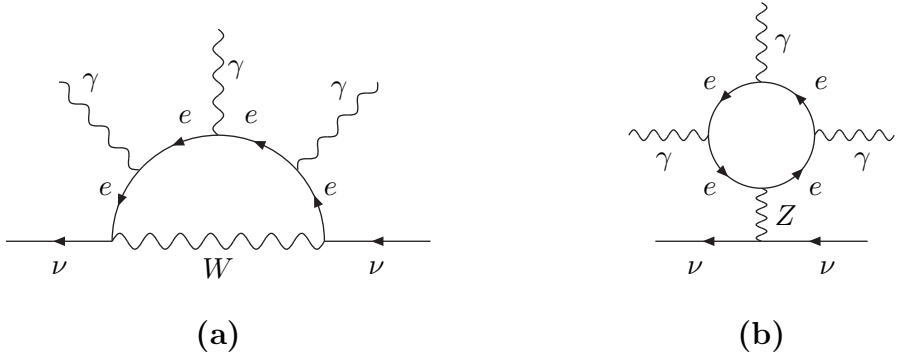
\begin{figure}[btp]
\begin{center}
%
%effective neutrino photon interaction
% 
\begin{picture}(180,120)(-90,-35)
\Text(0,-30)[ct]{\large\bf (a)}
\ArrowLine(80,0)(40,0)
\Text(60,-10)[c]{$\nu$}
\Photon(40,0)(-40,0)37
\Text(0,-10)[c]{$W$}
\ArrowLine(-40,0)(-80,0)
\Text(-60,-10)[c]{$\nu$}
%Changed
\ArrowArc(0,0)(40,0,45)
\Text(-15,47)[l]{$e$}
\ArrowArc(0,0)(40,45,90)
\Text(-40,20)[r]{$e$}
\ArrowArc(0,0)(40,90,135)
\Text(15,47)[r]{$e$}
\ArrowArc(0,0)(40,135,180)
\Text(45,20)[r]{$e$}
\Photon(-30,27)(-60,60){2}{4}
\Text(-45,60)[r]{$\gamma$}
\Photon(0,40)(0,80){2}{4}
\Text(10,65)[r]{$\gamma$}
%Changed
\Photon(30,27)(60,60){2}{4}
\Text(49,57)[r]{$\gamma$}
\end{picture}
%
%effective neutrino photon interaction
%
\qquad
\begin{picture}(180,120)(-90,-35)
\ArrowLine(40,0)(0,0) 
\Text(25,-10)[c]{$\nu$} 
\ArrowLine(0,0)(-40,0)
\Text(-25,-10)[c]{$\nu$} 
%\Photon(0,0)(0,30)24
\Photon(0,0)(0,20)24
\Text(5,10)[l]{$Z$}
\ArrowArc(0,40)(20,0,90)
\Text(20,60)[l]{$e$}
\Photon(0,60)(0,90)24
\Text(5,75)[l]{$\gamma$}
\ArrowArc(0,40)(20,90,180)
\Text(-22,60)[l]{$e$}
\Photon(-20,40)(-50,40)24
\Text(-40,30)[l]{$\gamma$}
\ArrowArc(0,40)(20,180,270)
\Text(-22,20)[l]{$e$}
\Photon(20,40)(50,40)24
\Text(35,30)[l]{$\gamma$}
\ArrowArc(0,40)(20,270,360)
\Text(18,20)[l]{$e$}
\Text(0,-30)[ct]{\large\bf (b)}
%\Text(0,-20)[]{\bf (b)}
\end{picture}
\caption[]{\sf The 1-loop effective vertex for two neutrinos and three
photons. 
\label{f:nugamma}}
\end{center}
\end{figure}
But there is no such restriction on the coupling of three photons with
neutrinos as,
\begin{eqnarray}
\gamma + \nu \to \gamma + \gamma + \nu.
%\label{}
\end{eqnarray}
The cross-section of the above process can be calculated from the
effective Lagrangian proposed by Dicus and Repko
\cite{Dicus:1997rw}. The diagrams for the two neutrino three photon
interaction are shown in Fig.~\ref{f:nugamma} where
Fig.~\ref{f:nugamma}a shows the contribution from the $W$ exchange
diagram and Fig.~\ref{f:nugamma}b shows the contribution from $Z$
exchange. Denoting the photon field tensor as $F_{\mu \nu}$ and the
neutrino fields by $\psi$, and integrating out the particles in the
loop the effective Lagrangian comes out as
\begin{eqnarray}
\mathscr L_{\rm eff} = {G_F \over \surd2}
\frac{e^3 (c_V+1)}{360 \pi^2 m^4}
\left[ 5 (N_{\mu \nu} F^{\mu \nu})(F_{\lambda \rho}
F^{\lambda \rho}) - 14 N_{\mu \nu} F^{\nu \lambda}
F_{\lambda \rho} F^{\rho \mu}\right],
\label{EL}
\end{eqnarray}
where $c_V$ was defined in \eqn{cvca}, and
\begin{eqnarray}
N_{\mu \nu} = \partial_\mu(\bar{\psi} \gamma_\nu L \psi) - 
\partial_\nu(\bar{\psi} \gamma_\mu L \psi) \,.
%\label{}
\end{eqnarray}

For energies much smaller than the electron mass, this can be used as
an effective Lagrangian to calculate various processes involving
photons and neutrinos in the presence of a background magnetic field
$B_{\mu\nu}$.  For this, we simply have to write
\begin{eqnarray}
F_{\mu\nu} = f_{\mu\nu} + B_{\mu\nu} \,,
%\label{}
\end{eqnarray}
where now $f_{\mu\nu}$ is the dynamical photon field, and look for the
terms involving $B_{\mu\nu}$.  For example, Shaisultanov
\cite{Shaisultanov:1997bc} calculated the rate of $\gamma\gamma\to
\nu\bar\nu$ in a background field.  \eqn{EL} shows that in the lowest
order, the amplitude for involving $\nu_e$'s would be proportional to
\begin{eqnarray}
{G_FB \over m_e^4} \sim {B \over M_W^2 m_e^2 B_e} \,,
%\label{}
\end{eqnarray}
where $B_e$ is the value of the magnetic field defined in \eqn{Bc}.
Since the amplitude without any magnetic field \cite{Dicus:iy} is of
order $1/M_W^4$, it follows that the background field increases the
amplitude by a factor of order $(M_W/m_e)^2 B/B_e$, or the rate by a
factor $(M_W/m_e)^4 (B/B_e)^2$.  Later calculations
\cite{Dicus:2000cz} have extended these results by including other
processes obtained by crossing, like $\nu\bar\nu\to\gamma\gamma$ and
$\nu\gamma\to\nu\gamma$.  To obtain higher $B$ terms in these cross
sections, one needs the effective Lagrangian containing higher order
terms in the electromagnetic field strength.  Such an effective
Lagrangian has been derived by Gies and
Shaisultanov~\cite{Gies:2000tc}.

Alternatively, the amplitudes can be calculated using the Schwinger
propagator for charged leptons.  Such calculations for
$\gamma\gamma\to \nu\bar\nu$ were done some time ago
\cite{Chyi:1999wy, Chyi:1999fc}.  One of the important features of
this calculation is that in the 4-fermi limit, the diagram contains
three electron propagators.  In such situations, the phase factor
$\Psi(x,x')$ appearing in the Schwinger propagator of
\eqn{schwingprop} cannot be disregarded.

In the calculation, only the linear term in $B$ was retained in the
amplitude so that the results are valid only for small magnetic
fields.  However, since no effective Lagrangian was used, the results
are valid even when the energies of the neutrinos and/or the photons
are comparable to, or greater than, the electron mass.  Later authors
\cite{Dicus:2000cz} reported some mistakes in this calculation and
corrected them.

%%%%%%%%%%%%%%%%%%%%
\section{Magnetic field effects on neutrino processes from external
charged particles}\label{ex}
%%%%%%%%%%%%%%%%%%%%
As discussed in Sec.~\ref{mo}, rates for neutrino processes are
modified in a background magnetic field because of the presence of
charged particles in the initial and/or final states.  Some such
processes might have very important astrophysical implications, some
of which will be discussed in Sec.~\ref{ef}.

%%%%%%%%%%%%%%%%%%%%
\subsection{Processes involving nucleons}\label{exnu}
%%%%%%%%%%%%%%%%%%%%
The charged current interaction Lagrangian involving neutrinos and
nucleons is given by
\begin{eqnarray}
\mathscr L_{\rm int} = \sqrt 2 \left[ \overline
\psi_{(e)} \gamma^\mu L \psi_{(\nu_e)} \right] \; 
\left[ \overline
\psi_{(p)} \gamma_\mu (G_V + G_A \gamma_5) \psi_{(n)} \right] \,,
\end{eqnarray}
where $G_V=G_F \cos\theta_C$, $\theta_C$ being the Cabibbo angle, and
$G_A/G_V=-1.26$.  This can be used to find the cross section for
various neutrino-nucleon scattering processes, as we describe now.

First we consider some processes in which a neutrino or an
antineutrino appears only in the final state.  In a star, when such
reactions occur, the final neutrino or the antineutrino escapes and 
the star loses energy.  Such processes are collectively known
as Urca processes, named after a casino in Rio de Janeiro where
customers lose money little by little \cite{Clayton}.  One such
process is the neutron beta-decay,
\begin{eqnarray}
n &\to& p + e^- +\bar\nu \,.
%\label{}
\end{eqnarray}
The rate of this process in a magnetic field was calculated by various
authors.  An early paper by Fassio-Canuto \cite{Can69} derived the
rate in a background of degenerate electrons.  Contemporary papers by
Matese and O'Connell \cite{MOc69,MOc70} derived the rate where the
background did not contain any matter, but included the effects of the
polarization of neutrons due to the magnetic field.  Protons and
neutrons were assumed to be non-relativistic in the calculations.
Further, the magnetic field was assumed to be much smaller than
$m_p^2/e$ so that its effect on the proton wave function could be
neglected.  The dispersion relation of \eqn{E} then suggests that the
Landau level of the electron is bounded by the relation
\begin{eqnarray}
N < {Q^2 - m_e^2 \over 2eB} \,,
%\label{}
\end{eqnarray}
where
\begin{eqnarray}
Q \equiv m_n - m_p \,.
%\label{}
\end{eqnarray}
The rate of the process should then include contribution from all
possible Landau levels in this range, and the general expression for
this rate was obtained \cite{MOc69}.

Other examples of Urca processes are
\begin{eqnarray}
p + e^- &\to& n + \nu \,,\\*
n + e^+ &\to& p + \bar\nu \,.   
%\label{}
\end{eqnarray}
The first reaction requires a threshold energy.  The second one is
possible at any energy.  Various calculations of these processes exist
in the literature.  Some calculations take the background matter
density into account \cite{Dorofeev:az, DRT85, Baiko:1998jq,
Bandyopadhyay:1998qs, Leinson:2001ei}, some include magnetic effects
on the proton wavefunction as well \cite{Leinson:2001ei}.

We now consider processes where neutrinos or antineutrinos appear in
the initial state only.  In a star, such processes contribute to the
opacity of neutrinos and antineutrinos.  One example of such process
is the inverse beta-decay process
\begin{eqnarray}
n + \nu &\to& p + e^- \,.
\label{invbeta}
\end{eqnarray}
The cross section of this process has been calculated by several
authors.  In the early calculation by Roulet \cite{Roulet:1997sw} and
by Lai and Qian \cite{Lai:1998sz}, the modification of the electron
wave function due to the magnetic field was not taken into account.
The magnetic field effects entered only through the following
modification of the phase space integral and the spin factor of the
electron:
\begin{eqnarray}
2\int {d^3p \over (2\pi)^3}  \longrightarrow {eB \over (2\pi)^2}
\sum_N g_N \int dp_z \,,
%\label{}
\end{eqnarray}
where $g_N$ is the degeneracy of the $N$-th Landau level, which is 1
for $N=0$ and 2 for all other levels.  The sum over $N$ is restricted
to the region
\begin{eqnarray}
N < {(Q+\omega)^2 - m_e^2 \over 2eB} \,,
%\label{}
\end{eqnarray}
where $\omega$ is the neutrino energy.

Subsequent calculations incorporated the modification of wave
functions.  Arras and Lai \cite{Arras:1998mv}, while still treating
the nucleons as non-relativistic, used the non-relativistic Landau
levels as well as the finiteness of the recoil energy for the proton.
They found the cross section and went on to derive expressions for the
neutrino opacity.  From the final expressions, one can only recognize
the terms linear in $B$.  The opacity was calculated also by Chandra,
Goyal and Goswami \cite{Chandra:2001at}.  Like the previous authors,
they also considered the contribution to the opacity from other
reactions like neutrino-nucleon elastic scattering.  In the work of
Bhattacharya and Pal \cite{Bhattacharya:1999bm, Bhattacharya:2002qf},
the cross section has been calculated ignoring nucleon recoil, but the
results are correct to all orders in $B$.

Incorporation of the proper wave functions of \eqn{Usoln} reveal a
property of the cross section that is not obtained by a mere
modification of the phase space.  The cross section is sensitive to
the angle $\theta$ between the neutrino momentum and the magnetic
field even if the neutron is unpolarized.  This is also true for the
Urca processes discussed above.  In addition, when one includes
neutron polarization, there are extra terms which depend on $\theta$.

%%%%%%%%%%%%%%%%%%%%
\subsection{Neutrino-electron scattering and related
processes}%\label{exel} 
%%%%%%%%%%%%%%%%%%%%
The cross-section for the elastic neutrino-electron scattering
\begin{eqnarray}
\nu + e \to \nu + e 
\label{nuenue}
\end{eqnarray}
was calculated by Bezchastnov and Haensel \cite{Bezchastnov:1996gy}
using the exact wave functions given in \eqn{Usoln}.  They considered
the reaction taking place in a background of electrons.

There are related processes, obtained by crossing, which contain a
neutrino-antineutrino pair in the final state.  For example, one can
have the pair annihilation of electron and positron into
neutrino-antineutrino: 
\begin{eqnarray}
e^- + e^+ \to \nu + \bar\nu \,.
%\label{}
\end{eqnarray}
In addition, one can consider the process
\begin{eqnarray}
e^- \to e^- + \nu + \bar\nu \,.
\label{eenunu}
\end{eqnarray}
This is similar to a synchrotron radiation reaction, the difference
being that a neutrino-antineutrino pair is produced instead of a
photon.  The process is usually called neutrino synchrotron radiation.
It should be noted that this process cannot occur in the vacuum.
However, in the presence of a background magnetic field and background
matter, the dispersion relation of the electron changes so that it
becomes kinematically feasible.  These processes provide important
mechanism for stellar energy loss, and the rates of these processes
have been calculated \cite{Kaminker:su, Kaminker:ic, Vidaurre:1995iv}.
The reverse of these processes are important for neutrino absorption,
and have also been studied~\cite{Hardy:2000gg}.

%%%%%%%%%%%%%%%%%%%%
\section{Possible implications}\label{ef}
%%%%%%%%%%%%%%%%%%%%
A magnetic field, as commented in Sec.~\ref{mo}, has sizeable effect
on neutrino properties if its magnitude is comparable to, or larger
than, $10^{13}\gauss$.  We also mentioned in Sec.~\ref{mo} that there
are stellar objects where such high fields presumably exist.  One can
therefore speculate the effects of such high fields on various
processes involving neutrinos on the equilibrium, dynamics and
evolution of these stars.

One of the effects of a strong magnetic field is to enhance the rate
of creation of neutrinos.  Various such processes were discussed in
Sec.~\ref{ex}, such as the Urca processes, $e^+$-$e^-$ pair
annihilation and neutrino synchrotron radiation.  We also commented
that, once produced, the neutrinos can easily go out of the star
because their mean free path is large.  Neutrino emission from young
neutron stars is the most important mechanism through which these
stars lose energy and become colder.  

Magnetic field enhances stellar energy loss in two ways.  First, as
the calculations show, the rates of different neutrino-producing
processes increase in the presence of a magnetic field.  Second,
processes such as the neutrino synchrotron radiation, \eqn{eenunu},
which cannot take place in the vacuum, become possible due to the
presence of a magnetic field and provide new channels for energy loss.

We now discuss processes like the neutrino-electron scattering and the
inverse beta-decay where neutrinos are not produced.  The first of
these processes, \eqn{nuenue}, controls the propagation of the
neutrinos from the core of the star to the boundary.  The second
process, \eqn{invbeta}, is directly related with the opacity of
neutrinos inside the star. The cross section of both these reactions
are enhanced in strong magnetic fields, implying that high magnetic
fields enhance not only the emissivity but also the opacity of
neutrinos.

A background magnetic field provides a preferred direction to a given
problem, and this shows up in the scattering cross-section of
neutrinos.  Enhancement and anisotropy of the cross-sections are
interrelated in a constant background magnetic field, but to make
things simpler we will discuss the two effects separately. The
anisotropic effects on the Urca processes have been calculated and
also we know how the reactions responsible for the opacity of
neutrinos respond to a unique direction of the magnetic field. There
is a specific aim for the calculations of anisotropic effects, viz.,
finding an explanation for the high velocities of the pulsars, of the
order of $450\pm90\;{\rm Km\,s}^{-1}$.  Typical pulsars have masses
between $1.0M_\odot$ and $1.5M_\odot$, i.e., about $2\times10^{33}$g.
The momentum associated with the proper motion of a pulsar would
therefore be of order $10^{41}$ g\,cm/s. On the other hand the energy
carried off by neutrinos in a supernova explosion is about
$3\times10^{53}$erg, which corresponds to a sum of magnitudes of
neutrino momenta of $10^{43}$g\,cm/s. Thus an asymmetry of order of
$1\%$ in the distribution of the outgoing neutrinos would explain the
kick of the pulsars. It has been argued that an asymmetry of this
order in the distribution of outgoing neutrinos can be generated by
the anisotropic cross-sections of the various neutrino related
processes in presence of a constant magnetic field \cite{Dorofeev:az,
Bisnovatyi-Kogan:1997an, Goyal:1998nq, Bhattacharya:1999bm}.  If, on
the other hand, the magnetic field is toroidal, anisotropic neutrino
emission can also produce a torque on the star and help regenerate the
magnetic field~\cite{Gvozdev:1999md}.

We next discuss some possible applications of the electromagnetic
interactions of neutrinos in a magnetic field background.  The photons
present in a neutron star are trapped due to their large
cross-sections with electrons or positrons.  Now
if due to the existence of a medium or of the magnetic field or of
both the dispersion relation of the photon is changed, real photons
can decay into neutrino-antineutrino pair, as discussed in
Sec.~\ref{viem}.  This provides new  ways of energy emission from the
star.  Other related processes involving photons and
neutrinos, such as
\begin{eqnarray}
\gamma + e^- &\to& e^- + \nu + \bar{\nu} \,,\\
\gamma + \gamma &\to& \nu + \bar{\nu} \,,
%\label{}
\end{eqnarray}
are also responsible for emission of neutrinos from the star.

Studies on propagation of neutrinos in a magnetic field has led to
investigations concerning their dispersion relation, as has been
discussed in Sec.~\ref{vise}.  Calculations of dispersion relation of
neutrinos has been done in vacuum and in a medium, for strong fields
and for weak fields.  An interesting cosmological consequence of these
dispersion relations has been discussed in the literature
\cite{Elizalde:2000vz}.  It is based upon the assumption that in the
time between the QCD phase transition epoch and the end of
nucleosynthesis, a cosmic magnetic field in the magnitude range
\begin{eqnarray}
m_e^2 \le eB \le M_W^2 
%\label{}
\end{eqnarray}
could have existed.  We have seen in \eqn{magdisp} that in presence of
a magnetic field, the neutrino dispersion relation acquires a
direction dependent term.  This would be reflected in the propagation
of neutrinos, and must leave its footprints in the neutrino relic
background.  Of course the effect would be appreciable only if $eB
\sim T^2$ where $T$ is the temperature during the neutrino decoupling
era, so that thermal fluctuations do not wash out these anisotropies.

Another interesting possible consequence of neutrino oscillations have
also been discussed \cite{Kusenko:1996sr} in the context of high
velocities of neutron stars.  In a material background containing
electrons but not any other charged leptons, the cross section of
$\nu_e$'s is greater than that of any other flavor of neutrino.  If
$\nu_e$'s can oscillate resonantly to any other flavor, they can
escape more easily from a star.  In a proto-neutron star, the resonant
density at an angle $\theta$ with the magnetic field occurs at a
distance $R_0+\delta \cos\theta$ from the center, where $\delta$ is a
function of the magnetic field and specifies the deformation from a
spherical surface.  So this distance is direction dependent, as we
have discussed in Sec.~\ref{vios}.  Therefore the escape of neutrinos
is also direction dependent, and the momentum carried away by them is
not isotropic.  The star would get a kick in the direction opposite to
the net momentum of escaped neutrinos.  This was suggested by Kusenko
and Segr\`e \cite{Kusenko:1996sr}, who estimated that the momentum
imbalance is proportional to $\delta$ and can have a magnitude of
around 1\% for reasonable values of $B$.  Later authors
\cite{Janka:1998kb} criticized their analysis and argued that the
effect was overestimated by them, because the kick momentum vanishes
in the lowest order in $\delta$.  A recent and detailed study
\cite{Barkovich:2002wh} indicates that these criticisms may not be
well-placed, and the kick momentum might indeed be proportional to the
surface deformation parameter $\delta$.

%%%%%%%%%%%%%%%%%%%%
\section{Concluding remarks} 
%%%%%%%%%%%%%%%%%%%%
Unfortunately, there is no conclusive remark on this subject.  Many
calculations have been done, but there is no consensus about whether
any of them explains any observational or experimental data.
If the magnetic field is small, the effect is small and cannot be
disentangled from the background.  Very large magnetic fields,
$B>B_e$, are obtained only at astronomical distances, presumably in
neutron stars and magnetars.  For such distant objects, observational
data are not clean enough to resolve the effects of the magnetic
field.  There are theses and anti-theses, but no synthesis so far.  The
calculations are in search of a physical effect to be explained by
them, much like the six characters in search of an author in the great
Italian dramatist Luigi Pirandello's play {\sl Sei personaggi in cerca
d'autore}.

%%%%%%%%%%%%%%%%%%%%
\paragraph*{Acknowledgments~: } 
%%%%%%%%%%%%%%%%%%%%
We have learned a great deal on the topics discussed in this short
review through various collaborations with Jos\'e F Nieves, Sushan
Konar and Avijit Ganguly.  We are indebted to them for what we know on
these issues.  We also thank Holger Gies for pointing out a careless
statement in an earlier version, which we have modified.

\end{document}